\documentclass[aps,prd,preprint,groupedaddress]{revtex4-1}

\usepackage{graphicx} 

\begin{document}

\title{ Thermal spectrum of pseudo-scalar  glueballs and Debye screening mass from holography  }



\author{Nelson R. F. Braga}\email{braga@if.ufrj.br}
\affiliation{Instituto de F\'{\i}sica,
Universidade Federal do Rio de Janeiro, Caixa Postal 68528, RJ
21941-972 -- Brazil}

\author{Luiz F.  Ferreira  }\email{luizfaulhaber@if.ufrj.br}
\affiliation{Instituto de F\'{\i}sica,
Universidade Federal do Rio de Janeiro, Caixa Postal 68528, RJ
21941-972 -- Brazil}

%\date{\today}

\begin{abstract} 
      
 The finite temperature spectrum of pseudo-scalar glueballs in a plasma is studied using a holographic model.  The $ 0^{-+}$ glueball is represented by a pseudo-scalar (axion) field living  in a five dimensional  geometry that comes from a solution of Einstein equations for  gravity   coupled with a dilaton scalar field.  
The spectral function obtained from the model shows a clear peak corresponding to the quasi-particle ground state. 
Analysing the variation of the position of the  peak with temperature, we describe the thermal behavior of the Debye screening mass of the plasma. 
As a check of consistency, the zero temperature limit of the model is also investigated. The  glueball masses obtained are consistent with previous  lattice results.

\end{abstract}

\keywords{Gauge-gravity correspondence, Phenomenological Models}

\maketitle

 \section{Introduction}
 
 The AdS/CFT correspondence \cite{Maldacena:1997re,Gubser:1998bc,Witten:1998qj} inspired  the development of holographic models that  describe strong interaction properties based on gauge/string duality.
Some of the first works  in this direction assumed the existence of an approximate duality  between a field theory living in some ad hoc deformation of anti-de Sitter (AdS) space containing a dimension-full parameter and a gauge theory where the parameter plays the role of an energy scale.
The simplest example is the  hard wall  AdS/QCD model,  that appeared in refs. \cite{Polchinski:2001tt,BoschiFilho:2002ta,BoschiFilho:2002vd}. It consists in placing a hard geometrical cutoff in AdS  space. 
This model provides, in a very simple way,  glueballs masses consistent with lattice results.
 Another AdS/QCD model, the soft wall, where  the square of the mass  grow linearly with the
radial excitation number, was introduced in ref. \cite{Karch:2006pv}. In this case, the background involves AdS space and a scalar field that acts effectively as a smooth  infrared cutoff.  
One finds an interesting review of AdS/QCD models and a wide list of  related references in \cite{Brodsky:2014yha}. 
  
 A finite temperature version of the AdS/CFT correspondence was found in  
 \cite{Witten:1998qj,Witten:1998zw}. In this case the gauge theory dual is a black hole geometry. 
 The corresponding finite temperature versions of AdS/QCD models provide a nice picture of the 
 confinement/deconfinement thermal phase transition\cite{BoschiFilho:2006pe,Herzog:2006ra,BallonBayona:2007vp}.
 
 The holographic approach to strong interaction has been widely  improved over the years.  In particular,  models  like 
 \cite{Gursoy:2007cb,Gursoy:2007er,Gursoy:2009jd,Gursoy:2008bu,Gursoy:2008za,Gubser:2008ny,Gubser:2008yx} that  use   supergravity backgrounds  coming from  consistent solutions of the Einstein equations,  mimic very  important results of QCD, like the running of the coupling constant.  These type of improved models require in general numerical solutions for the supergravity background. 
 
 Interesting analytical and numerical results for the thermal behavior of the plasma were
obtained also in \cite{Alanen:2009xs,Alanen:2009ej,Alanen:2009na,Alanen:2010tg,Kajantie:2011nx,Alanen:2011hh,Kajantie:2013gab}. For example, in ref. \cite{Kajantie:2011nx} the spectral function of the shear operator $T_{12}$ was calculated  in a hot Yang-Mills theory. 
  
 In ref. \cite{Li:2011hp} the thermodynamics of the plasma was studied using a simplified model 
 that has a first order phase transition. The results obtained for the sound speed, the entropy and other thermodynamic quantities are in good agreement with lattice results \cite{Boyd:1996bx},  showing that the model captures in a consistent way some important plasma properties.  
 
 The purpose of the present article is to apply the model of ref. \cite{Li:2011hp} to another important property of the plasma, namely,  the Debye screening mass. 
The approach that we will follow is to  study the thermal spectrum of pseudo-scalar $ 0^{-+}$  glueballs.  These particles are dual to the axion field in gauge gravity duality. 
 The mass of the ground state of $ 0^{-+}$  glueballs corresponds to the Debye screening mass of the plasma \cite{Arnold:1995bh,Bak:2007fk}.
 We will follow the prescriptions used in 
 refs. \cite{Gursoy:2007cb,Gursoy:2007er,Finazzo:2014zga,Finazzo:2015tta}
 for the action of an axion field. The spectral function obtained  presents a peak corresponding to the ground state of the pseudo scalar glueball.  
 So, the variation of the position of the peak with the temperature shows the thermal behavior of the  Debye screening mass. 
  
 The article is organized as follows. In section II we review the model of reference  \cite{Li:2011hp}. Then in section III we describe an axion field in this model and obtain the thermal spectrum of pseudo-scalar glueballs.  In section IV we study the zero temperature limit as a check of consistency of the model. Then we discuss the results obtained in the article in section V and analyze the temperature dependence of the Debye screening mass.

 \section{Holographic model with thermal phase transition }
 
The holographic bottom up model presented in ref. \cite{Li:2011hp} is constructed  using a 5-dimensional Einstein plus dilaton effective bulk action. The model is  a simplified version of 
the improved holographic QCD  (IHQCD) models of refs. \cite{Gursoy:2007cb,Gursoy:2007er}  that  presents  analytical solutions for the background.  

The effective five dimensional action for the metric and the dilaton in the Einstein frame is    
\begin{equation}\label{actionHQCD}
S=\frac{1}{16 \pi G_{5}}\int d^5x \sqrt{-g}\left[R-\frac{4}{3}(\partial \phi)^2-V_E(\phi) \right] \,,
\end{equation}
where $G_{5} $ is the five dimensional Newton's constant and  $ \phi $ the dilation field.  
The metric is assumed to have the form:
\begin{equation}\label{metric}
ds^2_{E}=\frac{e^{2A_{E}(z) }}{z^2} \left[-f(z)dt^2+d\vec{x}^2+\frac{dz^2}{f(z)}\right]\,,
\end{equation} 
\noindent  and the boundary is located at $z=0$. 

The potential $V_E(\phi )$  in the action (\ref{actionHQCD}) is not chosen a priori. The equations of motion relate $ A_E (z) $ , $\phi (z) $, $f(z) $  and $V_E (\phi) $.  The strategy to be followed is to choose a specific form for the 
warp factor and then determine the other quantities. 
It is convenient to replace the Einstein frame warp factor $A_E$ by the corresponding factor in the string frame $A_s$:
\begin{equation}
A_E (z) = A_{s}(z)-\frac{2}{3}\phi(z)\,.
\end{equation}

The equations of motion for the Einstein frame action (\ref{actionHQCD}) are 
\begin{eqnarray}
 \label{Einstein}
 && E_{\mu \nu}+ \frac{1}{2}g_{\mu \nu}\left( \frac{4}{3}\partial_{\mu} \phi \partial^\mu \phi+V_E(\phi) \right)-\frac{4}{3}\partial_{\mu} \phi \partial_{\nu} \phi=0 \,,\\
\cr
 \label{dilaton}
&&\frac{8}{3} \partial_z \left(\frac{e^{3A_s-2\phi}f}{z^3} \partial_z \phi \right)- \frac{e^{5A_s-\frac{10}{3}\phi}}{z^5}\partial_{\phi}V_E=0\, ,
\end{eqnarray}  
where the Einstein tensor is  $E_{\mu \nu}=R_{\mu\nu}-\frac{1}{2}g_{\mu \nu}R$ and the prime indicates differentiation with respect to z. 

The non-zero components of the gravity equation of motion (\ref{Einstein}) read
\begin{eqnarray}\label{Einsteintt}
&A_s''&+A_s'\left( \frac{f'}{2f}-\frac{2}{z}+A_s'-\frac{4}{3}\phi'\right)-f'\left(\frac{\phi'}{3f}+\frac{1}{2zf} \right)-\frac{2\phi''}{3}+\frac{2}{3}\phi' \left( \phi'+\frac{2}{z}\right)\cr &+&\frac{2}{z^2}+\frac{V_E(\phi)}{6z^2f}e^{2A_s-\frac{4}{3}\phi}=0 \,,
\end{eqnarray}

\begin{equation}\label{Einsteinzz}
\phi'^2-\phi' \left(4A_s'+\frac{f'}{2f}-\frac{4}{z} \right)+A_s'\left( \frac{3f'}{4f}-\frac{6}{z}+3A_s'\right)-\frac{3f'}{4zf}+\frac{3}{z^2}+\frac{V_E(\phi)}{4z^2f}e^{2As-\frac{4}{3}\phi}=0 \,,
\end{equation}

\begin{eqnarray}\label{Einsteinx1x1}
&f''&+f'\left( 6A_{s}'^2 -\frac{6}{z}  -4\phi' \right) + e^{2A_s-\frac{4}{3}\phi}\frac{V_E(\phi)}{z^2} \cr &+& f\left(6A_{s}''+6A_{s}'^{2}+\frac{12}{z^2}-4\phi''+4\phi'^2+\frac{8\phi'}{z} -\frac{4A_s'(2z\phi'+3)}{z}\right)=0 \,.
\end{eqnarray}  

Note  that  we only  need two of the above three equations.  The other equation  can be used as a consistency check for the  solutions. We recombine eqs.(\ref{Einsteintt}),  (\ref{Einsteinzz}) and  (\ref{Einsteinx1x1}) and find the following simplified equations:
\begin{equation}\label{eqsimp1}
f''+f'\left(3A_{s}'-2\phi'-\frac{3}{z}\right)=0\,,
\end{equation}  

\begin{equation}\label{eqsimp2}
\phi''-\phi'\left(2A_s'-\frac{2}{z}\right) -\frac{3A_s''}{2}-\frac{3A_s'}{z}+\frac{3}{2}A_{s}^2=0\,.
\end{equation}  
\noindent that determine the geometry once $A_s (z) $  is fixed. 
Integrating the previous equations one finds the solution in terms of $ A_s (z) $
 \begin{equation}\label{solphi}
\phi(z)=\phi_0+\frac{3A_{s}(z)}{2}+\phi_1\int^{z}_{0}\frac{e^{2A_s(x)}}{x^2}dx+\frac{3}{2}\int^{z}_{0}\frac{e^{2A_s(x)}\int^{x}_{0}y^2e^{-2A_s(y)}A_s'^2(y) \,dy \,}{x^2}dx\, ,
\end{equation}  

\begin{equation}\label{sol2}
f(z)=f_{0}+f_{1}\left(\int^{z}_{0}x^3e^{-3A_s(x)+2\phi(x)}dx\right) \,,
\end{equation}  
where $\phi_0$, $\phi_1$, $f_0$, $f_{1}$ are constants of integration. Note that $\phi (z)  $ in the second equation is obtained from the first equation, as a function of $A_s (z) $. 
The potential $V_E (\phi ) $ is also fixed by a choice 
of $A_s $. One just need to use the solutions  (\ref{solphi}) and (\ref{sol2})  in the equation:

\begin{equation}\label{sol3}
V_E(\phi)=e^{\frac{4}{3}\phi(z)-2A_s(z)}\left(z^2f''(z)-4f(z)(3z^2A_s''(z)-2z^2\phi''(z)+z^2\phi'^2(z)+3)\right)\,,
\end{equation}

The choice  used in ref.  \cite{Li:2011hp} for the  warp factor $A_s$   is:
\begin{equation}\label{dilaton}
A_s(z)=k^2z^2.
\end{equation}  
 
%Therefore,
%\begin{equation}\label{dilaton}
%A_E(z)=k^2z^2-\frac{2}{3}\phi(z) \,.
%\end{equation}  

 Imposing the asymptotic $AdS_5$ condition $f(0)=1$ near the $UV$ boundary $z \sim 0 $ and requiring $\phi$ and  $f$ to be finite at $z=0$ one finds
\begin{equation}\label{f}
f(z)=1-\frac{ \int^{kz}_0 x^3 \ exp \left(\frac{3}{2}x^2(H_c(x/k)-1)\right) }{\int^{kz_h}_0 x^3 \ exp \left(\frac{3}{2}x^2(H_c(x/k)-1)\right)} \,,
\end{equation} 

\begin{equation}\label{phi}
\phi(z)=\frac{3}{4}k^2z^2(1+H_c(z))\,,
\end{equation}
where:
\begin{equation}\label{Hc}
H_c(z)=\ _{2}F_{2}\left(1,1;2,\frac{5}{2};k^2z^2 \right) \,. 
\end{equation}
The solution for the dilaton potential take the form as following:

\begin{eqnarray}
V_E(z)&=&\frac{3e^{k^2z^2(1-H_c(z))}}{128k^2z^2}f(z)[40k^2z^2+64k^4z^4-384k^6z^6-27\pi e^{4k^2z^2}Er f_c(\sqrt{2}kz)^2\nonumber\\&&+12\sqrt{12} e^{2k^2z^2}kz(-7+20k^2z^2)Er f_c(\sqrt{2}kz)]-\frac{3f^h e^{\frac{5}{2}k^2z^2(-1+H_{c}(z))}k^3z^3}{16}\nonumber\\&&[4kz-16k^3z^3+3\sqrt{2\pi}e^{2k^2z^2}Er f_c(\sqrt{2}kz)]\, ,
\end{eqnarray}\\
where  $Er f_c[z]$  is error function  which is defined as a integral form  $Er f[z]=\frac{2}{\sqrt{\pi}}\int_{0}^{z}e^{-t^2}dt$,  $f^{h}$ is
\begin{equation}\label{fh}
f^{h}=\frac{1}{\int^{kz_h}_0 x^3 \ exp \left(\frac{3}{2}x^2(H_c(x/k)-1)\right)}\,.
\end{equation}

The temperature    is   obtained from:
\begin{equation}\label{T1}
T=\frac{|f'(z_h)|}{4 \pi}\,.
\end{equation}
Using eq.(\ref{f}),  one can find the relation between the temperature and the position of the black hole horizon in this model

\begin{equation}\label{T2}
T(z_h)=\frac{k^4z_{h}^3 \ exp \left(\frac{3}{2}(k^2z_{h}^2H_c(z_h)-k^2z_{h}^2)\right)}{4 \pi \int^{kz_h}_0 x^3 \ exp \left(\frac{3}{2}x^2(H_c(x/k)-1)\right) }.
\end{equation} 

The entropy is given by 
\begin{equation}\label{entropy}
s=\frac{A_{area}}{4G_{5}V_3}\bigg|_{z_h}=\frac{L^3}{4G_{5}}\left(\frac{e^{A_{E}(z)}}{z}\right)^3\bigg|_{z_h}
\end{equation}

Using $k=0.43 $  GeV and $G_5/L^3=1.26$ as in  ref. \cite{Li:2011hp}  one obtains the numerical results  for the temperature and entropy shown in Fig.(1).
  \\
  
 \begin{figure}[h]
\label{g6}
\begin{center}
\includegraphics[scale=0.4]{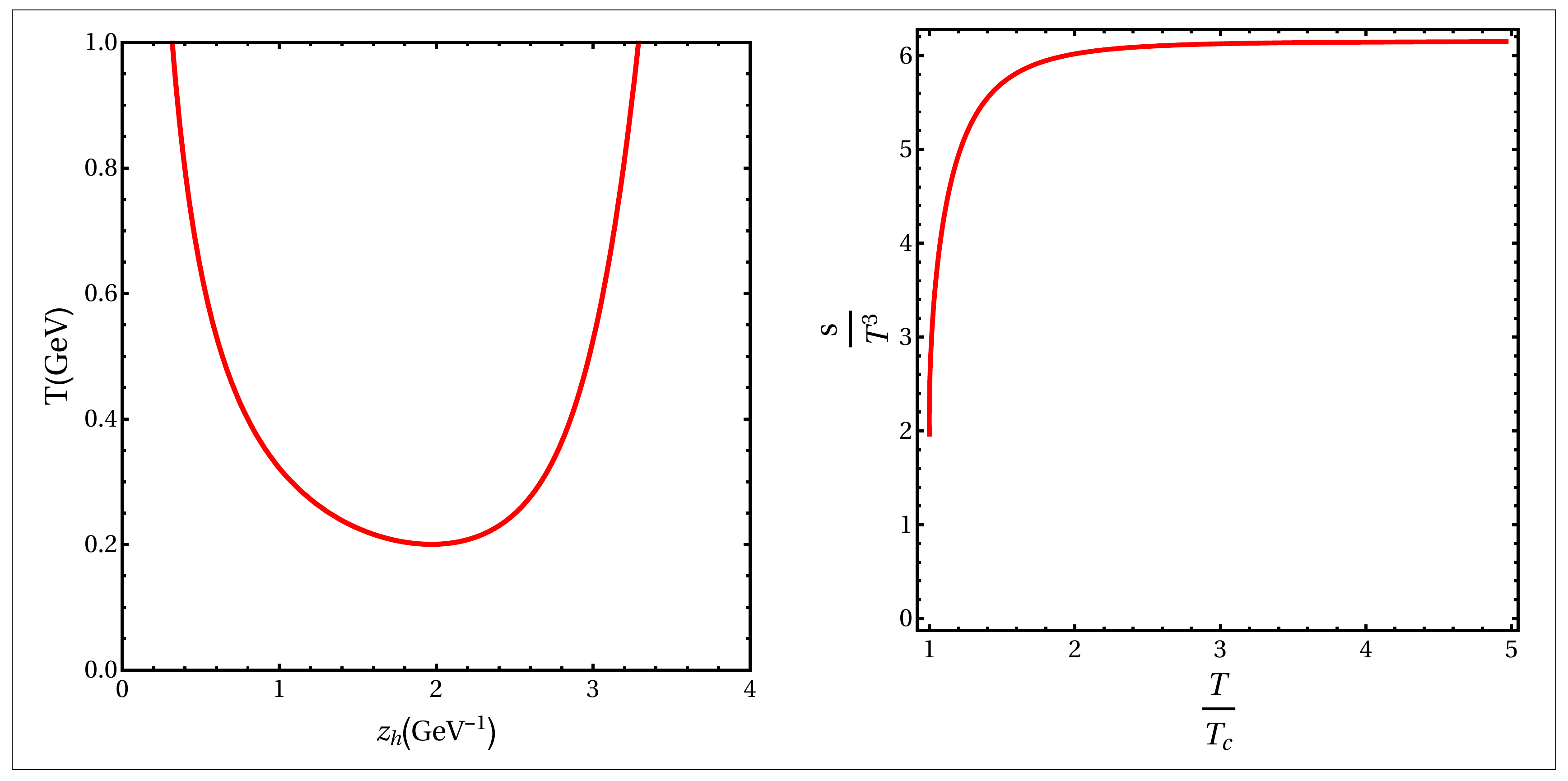}
\end{center}
\caption{The left panel shows the temperature $T$ as a function of the black-hole horizon $z_h$ with $k=0.43  $ GeV. In the right panel we present the scaled entropy density $s/T^3$ as a function of scaled temperature $T/T_{c}$ with $k=0.43  $ GeV and $G_5/L^3=1.26$.}
\end{figure}

 Note from Fig.(1)  that there is a minimal temperature $T_{min}$ at a certain black hole horizon position $z_{h}^m$. For $T<T_{min}$, there are no black hole solutions while for  $T>T_{min}$, there are two black hole solutions. When $z_h<z_{h}^m$, the temperature increase with the decrease of $z_h$, this phase is thermodynamically stable. When $z_h>z_{h}^m$, the temperature increases with the increase of $z_h$, this phase is thermodynamically unstable and thus not physical.

 \begin{figure}[h]
\label{g6}
\begin{center}
\includegraphics[scale=0.5]{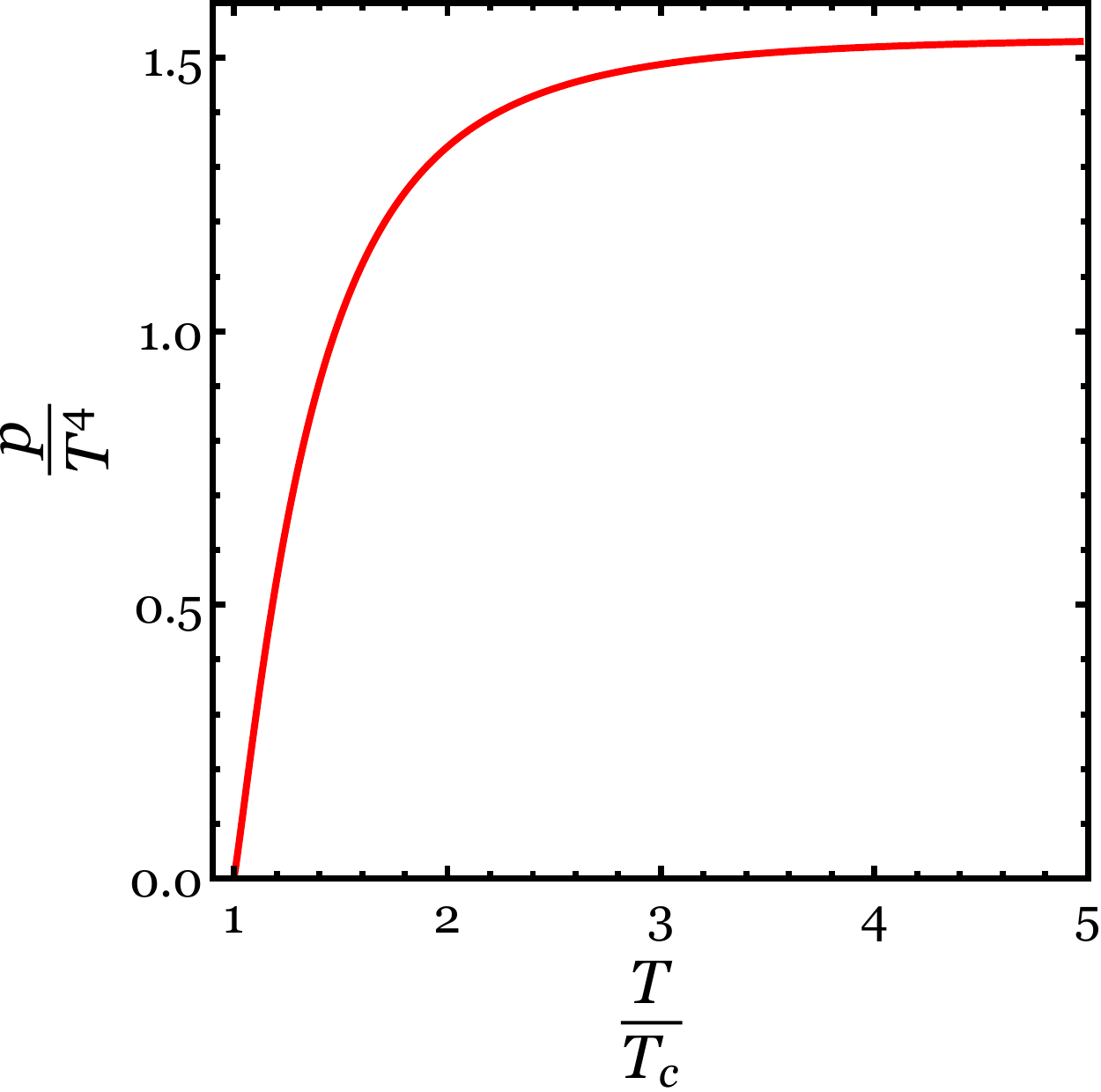}
\end{center}
\caption{The scaled pressure density $p/T^4$ as a function of scaled temperature $T/T_{c}$ with $k=0.43 $  GeV and $G_5/L^3=1.26$. }
\end{figure}
The pressure density $p(T)$ can be calculated from the entropy density  $s(T)$ by solving the equation:
\begin{equation}\label{pressure}
\frac{dp(T)}{dT}=-s(T)\,.
\end{equation}

After integrating  eq.(\ref{pressure}), the pressure density of the system can be obtained up to a integral constant $p_0$. One can set $p_0=0$ to ensure that $p(T_{min})=0$. Thus, the critical temperature  of this  model is in $T_c=T_{min}=201 MeV$. In  Fig.(2), we present the numerical result of pressure density as a function of temperature.
  
The sound velocity $c_{s}$ can be derived from the temperature  and entropy:
\begin{equation}\label{soundvelocity}
c^2_{s}=\frac{d  \log T}{d  \log s}\,.
\end{equation}
From  eq.(\ref{soundvelocity}) one  can see that the sound velocity is independent of the $5D$ Newton constant $G_5$. In Fig.(3), we show the  equation of state for this model.
 \begin{figure}[h]
\label{g6}
\begin{center}
\includegraphics[scale=0.5]{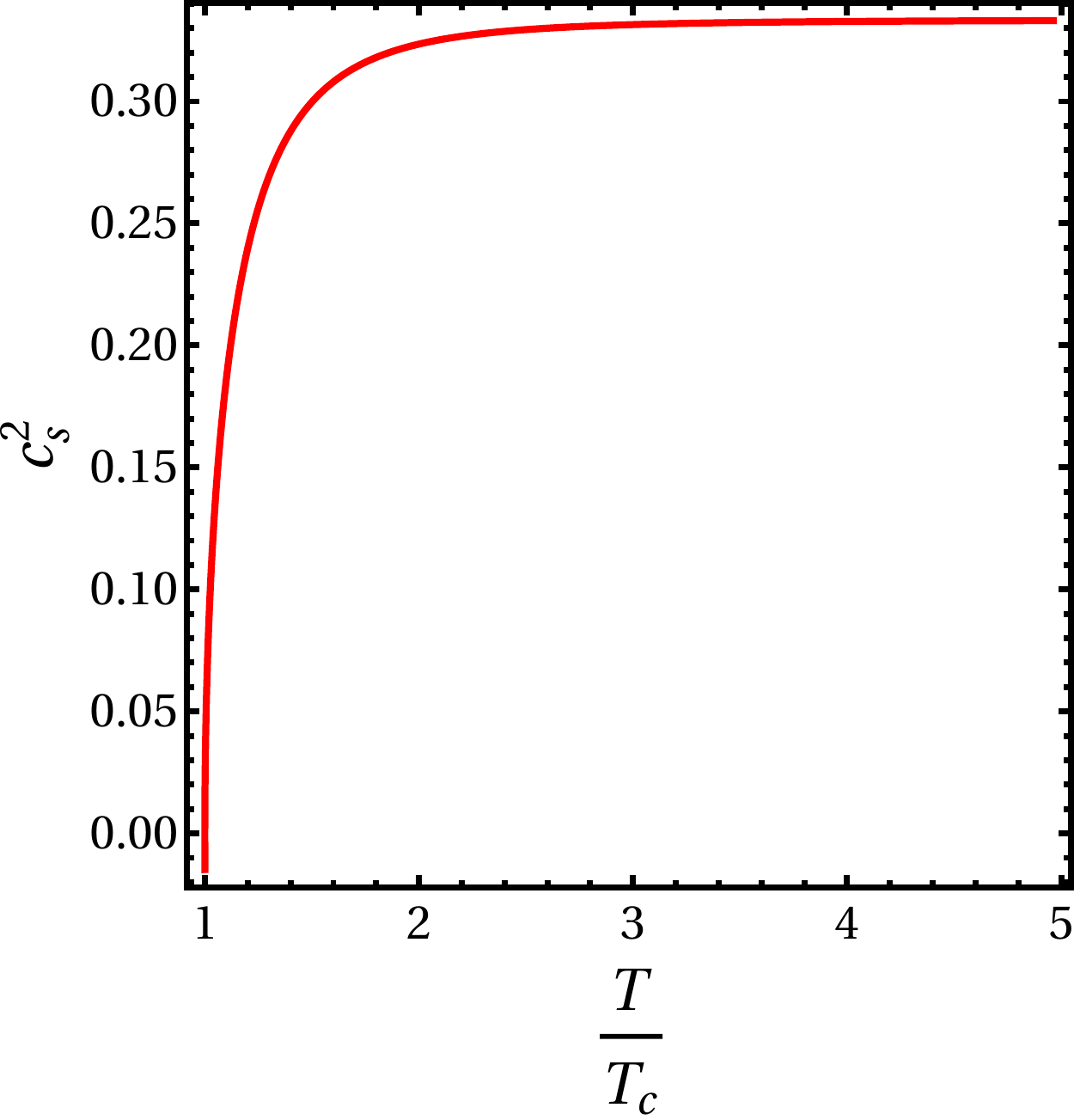}
\end{center}
\caption{The square of the sound velocity $c^2_{s}$ as a function of scaled temperature  $T/T_{c}$ with $k=0.43 $ GeV and $G_5/L^3=1.26$. }
\end{figure}

\section{Debye Mass from Pseudo-scalar glueballs }

\subsection{ Debye screening mass} 

An important concept in the description of the deconfined phase of non Abelian gauge theories at finite temperature is the Debye Screening mass $m_D$. The inverse of this quantity can be used to define a screening length felt by color-electric  excitations, in analogy with the Debye screening in Abelian plasma, which is felt by electric fields but not by magnetic fields.  

A gauge invariant non-perturbative definition for $m_D$ was given in ref. \cite{Arnold:1995bh} where  this quantity is defined as the smallest inverse correlation length in symmetry channels which are odd under Euclidean time reflection. 

 In this work we follow 
\cite{Bak:2007fk} and identify the smallest thermal mass,  associated with the Pontryagin density operator $ Tr ( F_{\mu \nu}\tilde{F}^{\mu \nu} ) $, as the Debye mass in a strongly coupled plasma. Thus, we can use the spectral function of the pseudo-scalar glueball to find the Debye screening mass  of the plasma. The supergravity field associated with the $ 0^{-+}$ operator is the axion.

\subsection{ Spectral function for pseudo-scalar glueballs} 

In order to describe the $ 0^{-+}$  glueball we consider the axion field in the supergravity background presented in the previous section.   
The action for the massless axion fluctuation, $ a $,  is assumed to be of the same  form as in Improved Holographic QCD \cite{{Gursoy:2007cb},{Gursoy:2007er},{Gursoy:2009jd},{Gursoy:2008bu},{Gursoy:2008za}}

\begin{equation}\label{acaoaxion}
S=\frac{1}{32 \pi G_{5}}\int  d^5x \sqrt{g}\mathcal{Z}(z)g^{\mu \nu}\partial_{\mu}a \partial_{\nu}a\, ,
\end{equation}
where  the axion coupling $\mathcal{Z}$  is a function which represents a partial resummation of high orders forms coming from string theory \cite{{Gursoy:2007cb},{Gursoy:2007er}}.  The convenient parametrization for the axion coupling found in these references is
\begin{equation}\label{Zparame}
\mathcal{Z} =1+c \lambda^4\,,
\end{equation}
where $c$ is a constant and $\lambda= \lambda(z) = e^{\phi (z) }$ is the 't Hooft coupling.
 
The equation of motion that come from action (\ref{acaoaxion}) with metric (\ref{metric})  is
\begin{equation}\label{eqmotion}
\partial_{\mu}\left( \, \mathcal{Z} \, b^5 g^{\mu \nu}\partial_{\nu}  a \right) \, = \, 0\,,
\end{equation}
where $b(z)=e^{A_{E}(z)}/z$. In Fourier space eq.(\ref{eqmotion}) reads :
\begin{equation}\label{eqtosolve}
\partial_z \left( \mathcal{Z} \, b^3 \, f \, \partial_z a \right)+\left( \frac{\omega^2}{f}-\vec{k}^2\right)\mathcal{Z} \, b^3 \, a=0 \,.
\end{equation}

In order to obtain the spectral function  of the pseudo-scalar glueball we follow the procedure of the membrame paradigm \cite{Iqbal:2008by}  as explained in ref. \cite{Gursoy:2012bt}. 
We first introduce the bulk response function 
\begin{equation}\label{responsebulk}
\xi(z,\omega,\vec{k})\equiv\frac{\Pi(z,\omega,\vec{k})}{\omega a(z,\omega,\vec{k})}\,,
\end{equation}
where $\Pi(z,\omega,\vec{k})$ is the radial canonical  momentum conjugate to the axion fluctuation $a(z,\omega,\vec{k})$:
\begin{equation}\label{momentum}
\Pi(z,\omega,\vec{k})\equiv \frac{\delta S}{\delta(\partial_z a) }=-\mathcal{Z} \,b^3 \, f \, \partial_z a(z,\omega,\vec{k})\,.
\end{equation}
This procedure allows us to reduce the linear second order differential equation (\ref{eqtosolve}) to a first order nonlinear equation:
\begin{equation}\label{eqmembrane}
\partial_{z}\xi-\frac{\omega}{\mathcal{Z} \, b^{3} \, f }\left[ \xi^2+\mathcal{Z}^2 \,  b^{6} \left(1- \, f \,\frac{\vec{k}^2}{\omega^2}\right)\right]=0\,.
\end{equation}
Requiring regularity at the horizon, one obtains the following horizon condition, needed to solve the first order equation above:
\begin{equation}\label{horizoncondition}
\xi(z_{h})=i \left[ \mathcal{Z} \, b^3 \right]_{ z = z_h }.
\end{equation}
From the Kubo's formula for the retarded Green's function we have the following relation:
\begin{equation}\label{Greenfunction}
G_{R}(\omega,\vec{k})=-\lim_{z\rightarrow0}\frac{\Pi(z,\omega,\vec{k})}{ a(z,\omega,\vec{k})}=-\omega \lim_{z\rightarrow0}\xi(z,\omega,\vec{k}).
\end{equation}

Thus, the spectral function is obtained  solving the equation (\ref{eqmembrane}) and using the imaginary part of the Green's function (\ref{Greenfunction}): 
\begin{equation}\label{spectral}
\rho(\omega)=-2 \ Im \ G_{R}(\omega,\vec{k}=0).
\end{equation}

\subsection{ Results}
 
 We fix  the value of the parameter in eq. ({\ref{Zparame}) as $c=0.26$ in order to 
 find a zero temperature limit  for  the mass of the pseudo-scalar glueball consistent  
 with the lattice results, as discussed in refs.   \cite{Gursoy:2007cb,Gursoy:2007er}. 
 We also use $ k$ = 0.43 GeV. 
 
%%%%%%%%%%%%%%%%%%%

\begin{figure}[h]
\label{g67}
\begin{center}
\includegraphics[scale=0.6]{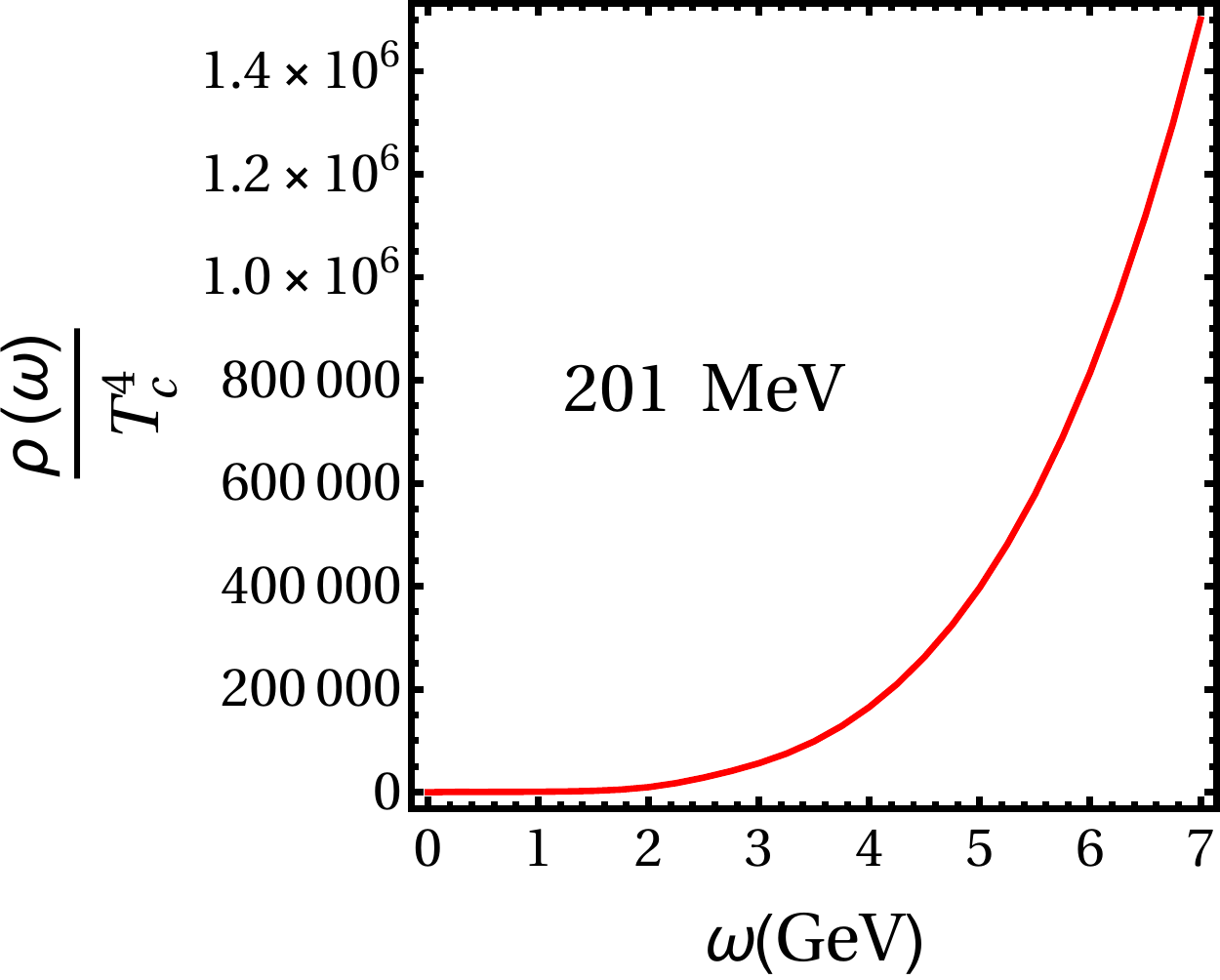}
\end{center}
\caption{Spectral function divided by $T_c^4 $ as a function of the energy for the critical temperature. }
\end{figure}

 The next step is  to solve  numerically eq.(\ref{eqmembrane}) using the metric  (\ref{metric})  and
 the axion coupling  (\ref{Zparame})  to find the spectral function  (\ref{spectral}). 
 We show in   Fig. (4)  the spectral function for the case  $T =T_c = 201 $  MeV.
 Note that we divide the spectral  function by $T_c^4$ to have a dimensionless quantity. 
 The numerical results show that for large $\omega$  the spectral function scales  as $\omega^4$. 
So, we  define a re-scaled  spectral function, in a similar way as done is refs. 
\cite{Grigoryan:2010pj,Braga:2016wkm}:
 \begin{equation}\label{ScaledSpec}
\tilde{\rho}(\omega)=\frac{\rho(\omega)}{\omega^4}\,.
\end{equation}
The results obtained  for   $\tilde{\rho}$ from numerical calculations at different  temperatures are shown in Fig.(5).  The location of the peak corresponds to the  thermal mass of the ground state. Increasing the temperature of the plasma one can observe in Fig.(5) that the peak decreases and virtually disappears for temperatures greater than  $T=230 \ $MeV. 

%%%%%%%%%%%%%%%%%

\begin{figure}[h]
\label{g6}
\begin{center}
\includegraphics[scale=0.35]{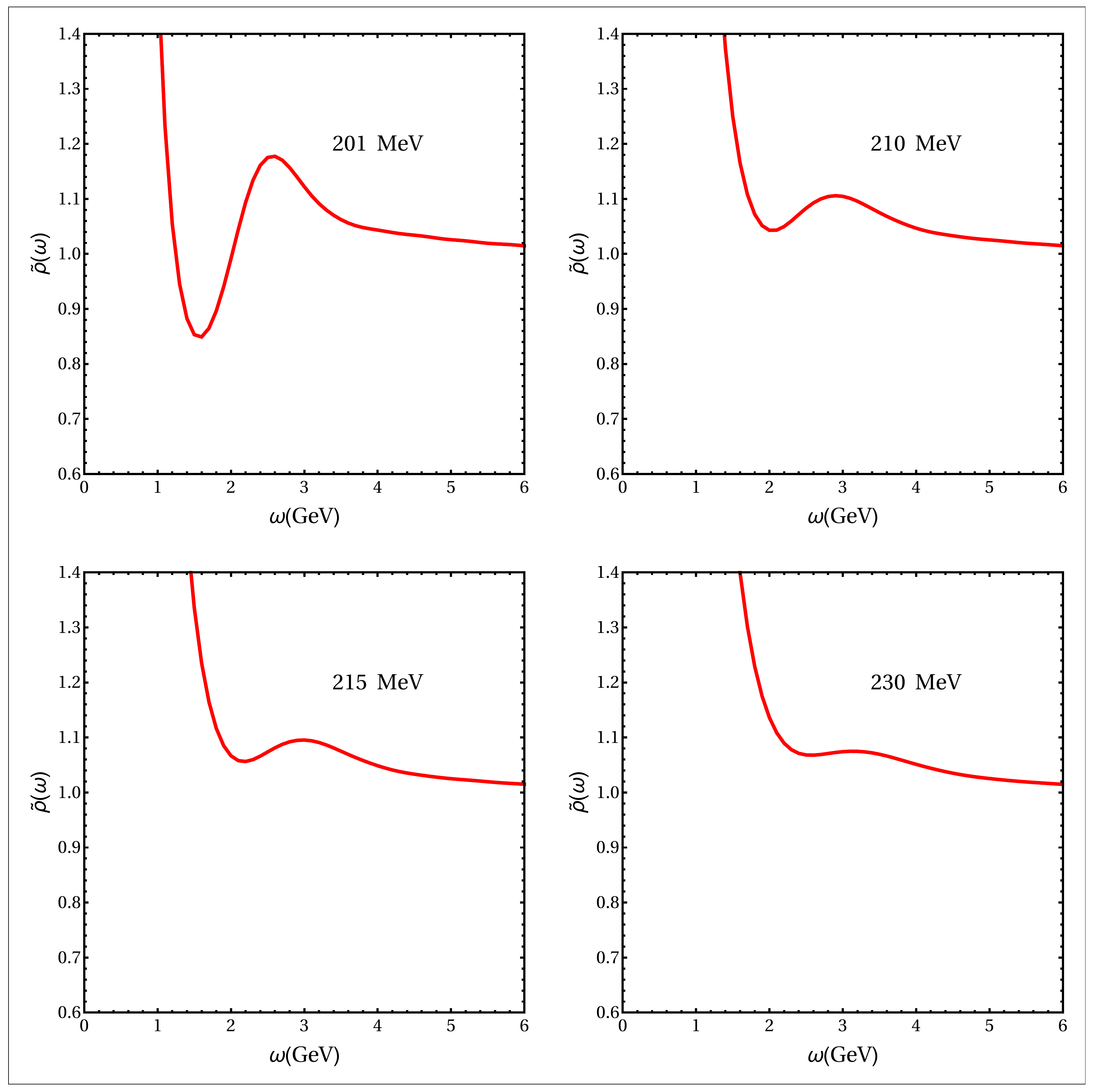}
\end{center}
\caption{Rescaled spectral function as a function of the energy $\omega$ in GeV }
\end{figure}
  
%%%%%%%%%%%%%%%%%%%

%\newpage

\section{Pseudo-scalar glueball  masses at  $ T $ =0}
 As a check of the method used for describing pseudo-scalar glueballs inside a plasma, let us 
 consider the limit of zero temperature. 
  The equation of motion for the axion is (\ref{eqtosolve})
    \begin{equation}\label{eqtosolve1}
\partial_z \left( \mathcal{Z} \, b^3 \, f \, \partial_z a \right)+\left( \frac{\omega^2}{f}-\vec{k}^2\right)\mathcal{Z} \, b^3 \, a=0 \,.
\end{equation}
 In order to calculate the glueball $0^{-+}$ masses at  $T=0$  one takes  $f(z)= 1$ and equation (\ref{eqtosolve1}) becomes:
  \begin{equation}\label{eqtosolve2}
\partial_z \left( \mathcal{Z} \, b^3  \, \partial_z a \right)+\left( \omega^2  -\vec{k}^2\right)\mathcal{Z} \, b^3 \, a=0 \,.
\end{equation}
 Defining  $\psi=e^{B}a$ and  
\begin{equation}\label{eqB}
B(z)=\frac{3}{2}\log b(z)+\frac{1}{2}\log \mathcal{Z}(z)\,
\end{equation}
the equation of motion (\ref{eqtosolve2})  takes the form 
\begin{equation}\label{eqSch}
-\psi''(z)+\mathcal{V}(z)=M^2\psi,
\end{equation}
where  $M^2=\omega^2-\vec{k}^2$ and the potential is defined as
\begin{equation}\label{eqPot}
\mathcal{V}(z)=B'^{\,2}+B''.
\end{equation}

Now, we can calculate the glueball mass using eq.  (\ref{eqSch}) and the metric (\ref{metric}) with $f(z)=1$.  The form of $A_E$ and $\phi$ is the same as in the finite temperature case. 

The parametrization for the axion coupling $ \mathcal{Z}(\lambda(z)) $ is the same as in 
eq. (\ref{Zparame}) with the same choice $c=0.26$ as in the finite temperature case of the previous section.  We present the results for the masses of $0^{-+}$ glueballs  in  Table (1),  comparing with lattice results \cite{{Morningstar:1999rf},{Chen:2005mg}}.  One notes that there is a reasonable agreement.

\begin{table}[h]
\centering
\vspace{0.5cm}
\begin{tabular}{|l|c|c|c|}\hline
$J^{PC}$ & Holographic Mass  \ &  Lattice Mass \cite{Morningstar:1999rf}&  Lattice Mass\cite{Chen:2005mg} \\ % Note a separação de col. e a quebra de linhas
\hline                               % para uma linha horizontal
$0^{-+}$ & 2.477  (GeV)       & 2.590 (GeV) & 2.560 (GeV)\\\hline
$0^{*-+}$ &  3.617 (GeV) &  3.640 (GeV)& \\\hline
$0^{**-+}$  & 4.630   (GeV)       & & \\\hline
\end{tabular}
\caption{Our holographic results for Glueball masses at zero temperature, with $ k \,  =\,$ 0.43 GeV, compared with lattice data \cite{Morningstar:1999rf,Chen:2005mg}.}
\end{table}

\section{ Analysis of the results } 

The location of the peak of the spectral functions obtained in section III corresponds to the thermal mass of the ground state of the  pseudo-scalar glueball, associated with the Debye screening mass of the plasma.

%%%%%%%%%%%%%%%%%%%%%%
\begin{figure}[h]
\label{g6}
\begin{center}
\includegraphics[scale=0.7]{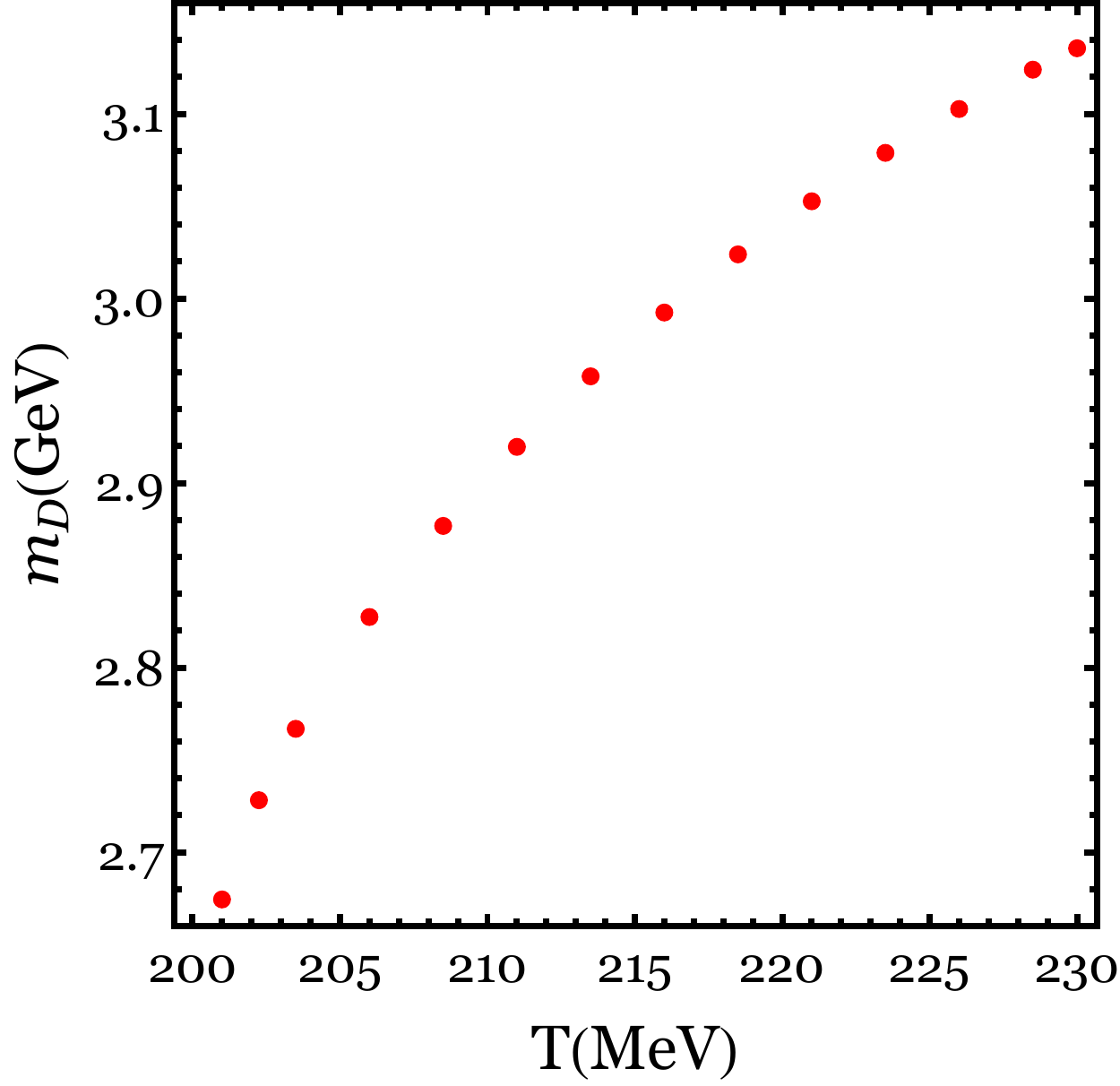}
\end{center}
\caption{Debye mass in GeV as a function of the temperature in MeV  }
\end{figure}
%%%%%%%%%%%%%%%%%%%%%%%%%

In order to exhibit  the thermal behavior of the Debye mass we plot in Fig.(6)  some values of the peak locations at different temperatures.  The smallest value of the mass occurs at the critical temperature $T_c$.  One notes that  for increasing temperature, the
plasma becomes more and more screened since  $m_D$ is monotonically increasing. 
The model provides the values of the  screening mass in the range:  201  MeV  $ \le  T \le   $ 230 MeV,
making it possible to investigate the region near the critical temperature.  

An important support for the validity of the model comes from the check, in the previous section, of the   $T = 0$ limit. The value obtained for the glueball mass is consistent with lattice data.  
Other studies of screening in a non Abelian plasma using holography can be found,  for example,  in 
\cite{Bak:2007fk,Finazzo:2014zga,Hoyos:2011uh,Singh:2012xj,Andreev:2016hxm}.

\noindent {\bf Acknowledgments:}   N.B. is partially supported by CNPq (Brazil) and L. F.  is supported by CAPES (Brazil).


\begin{thebibliography}{9}


\bibitem{Maldacena:1997re}
  J.~M.~Maldacena,
  %``The large N limit of superconformal field theories and supergravity,''
  Adv.\ Theor.\ Math.\ Phys.\  {\bf 2}, 231 (1998)
  [Int.\ J.\ Theor.\ Phys.\  {\bf 38}, 1113 (1999)].
 [arXiv:hep-th/9711200].
  %%CITATION = HEP-TH 9711200;%%

%\cite{Gubser:1998bc}
\bibitem{Gubser:1998bc}
  S.~S.~Gubser, I.~R.~Klebanov and A.~M.~Polyakov,
  %``Gauge theory correlators from non-critical string theory,''
  Phys.\ Lett.\  B {\bf 428}, 105 (1998).
  [arXiv:hep-th/9802109].
  %%CITATION = PHLTA,B428,105;%%

%\cite{Witten:1998qj}
\bibitem{Witten:1998qj}
  E.~Witten,
  %``Anti-de Sitter space and holography,''
  Adv.\ Theor.\ Math.\ Phys.\  {\bf 2}, 253 (1998).
  [arXiv:hep-th/9802150].
  %%CITATION = 00203,2,253;%%
  

  
  %\cite{Polchinski:2001tt}
\bibitem{Polchinski:2001tt}
  J.~Polchinski and M.~J.~Strassler,
  %``Hard scattering and gauge/string duality,''
  Phys.\ Rev.\ Lett.\  {\bf 88}, 031601 (2002)
  [arXiv:hep-th/0109174].
  %%CITATION = PRLTA,88,031601;%%

 %\cite{BoschiFilho:2002ta}
\bibitem{BoschiFilho:2002ta}
  H.~Boschi-Filho and N.~R.~F.~Braga,
  %``QCD / string holographic mapping and glueball mass spectrum,''
  Eur.\ Phys.\ J.\  C {\bf 32}, 529 (2004)
  [arXiv:hep-th/0209080].
  %%CITATION = EPHJA,C32,529;%%
  
%\cite{BoschiFilho:2002vd}
\bibitem{BoschiFilho:2002vd}
  H.~Boschi-Filho and N.~R.~F.~Braga,
  %``Gauge / string duality and scalar glueball mass ratios,''
  JHEP {\bf 0305}, 009 (2003)
  [arXiv:hep-th/0212207].
  %%CITATION = JHEPA,0305,009;%%
  
   
  %\cite{Karch:2006pv}
\bibitem{Karch:2006pv} 
  A.~Karch, E.~Katz, D.~T.~Son and M.~A.~Stephanov,
  %``Linear confinement and AdS/QCD,''
  Phys.\ Rev.\ D {\bf 74}, 015005 (2006)
  doi:10.1103/PhysRevD.74.015005
  [hep-ph/0602229].
  

  %\cite{Brodsky:2014yha}
\bibitem{Brodsky:2014yha} 
  S.~J.~Brodsky, G.~F.~de Teramond, H.~G.~Dosch and J.~Erlich,
  %``Light-Front Holographic QCD and Emerging Confinement,''
  Phys.\ Rept.\  {\bf 584}, 1 (2015)
  doi:10.1016/j.physrep.2015.05.001
  [arXiv:1407.8131 [hep-ph]].
   

 %\cite{Witten:1998zw}
\bibitem{Witten:1998zw} 
  E.~Witten,
  %``Anti-de Sitter space, thermal phase transition, and confinement in gauge theories,''
  Adv.\ Theor.\ Math.\ Phys.\  {\bf 2}, 505 (1998)
  [hep-th/9803131].
 
%\cite{BoschiFilho:2006pe}
\bibitem{BoschiFilho:2006pe} 
  H.~Boschi-Filho, N.~R.~F.~Braga and C.~N.~Ferreira,
  %``Heavy quark potential at finite temperature from gauge/string duality,''
  Phys.\ Rev.\ D {\bf 74}, 086001 (2006)
  doi:10.1103/PhysRevD.74.086001
  [hep-th/0607038].

  %\cite{Herzog:2006ra}
\bibitem{Herzog:2006ra} 
  C.~P.~Herzog,
  %``A Holographic Prediction of the Deconfinement Temperature,''
  Phys.\ Rev.\ Lett.\  {\bf 98}, 091601 (2007)
  doi:10.1103/PhysRevLett.98.091601
  [hep-th/0608151].
  
  %\cite{BallonBayona:2007vp}
\bibitem{BallonBayona:2007vp} 
  C.~A.~Ballon Bayona, H.~Boschi-Filho, N.~R.~F.~Braga and L.~A.~Pando Zayas,
  %``On a Holographic Model for Confinement/Deconfinement,''
  Phys.\ Rev.\ D {\bf 77}, 046002 (2008)
  doi:10.1103/PhysRevD.77.046002
  [arXiv:0705.1529 [hep-th]].
   

%\cite{Gursoy:2007cb}
\bibitem{Gursoy:2007cb} 
  U.~Gursoy and E.~Kiritsis,
  %``Exploring improved holographic theories for QCD: Part I,''
  JHEP {\bf 0802}, 032 (2008)
  doi:10.1088/1126-6708/2008/02/032
  [arXiv:0707.1324 [hep-th]].
  %%CITATION = doi:10.1088/1126-6708/2008/02/032;%%
  %280 citations counted in INSPIRE as of 06 Feb 2017
  
  
 %\cite{Gursoy:2007er}
\bibitem{Gursoy:2007er} 
  U.~Gursoy, E.~Kiritsis and F.~Nitti,
  %``Exploring improved holographic theories for QCD: Part II,''
  JHEP {\bf 0802}, 019 (2008)
  doi:10.1088/1126-6708/2008/02/019
  [arXiv:0707.1349 [hep-th]].
  %%CITATION = doi:10.1088/1126-6708/2008/02/019;%%
  %288 citations counted in INSPIRE as of 06 Feb 2017
 
%\cite{Gursoy:2009jd}
\bibitem{Gursoy:2009jd} 
  U.~Gursoy, E.~Kiritsis, L.~Mazzanti and F.~Nitti,
  %``Improved Holographic Yang-Mills at Finite Temperature: Comparison with Data,''
  Nucl.\ Phys.\ B {\bf 820}, 148 (2009)
  doi:10.1016/j.nuclphysb.2009.05.017
  [arXiv:0903.2859 [hep-th]].
  %%CITATION = doi:10.1016/j.nuclphysb.2009.05.017;%%
  %110 citations counted in INSPIRE as of 06 Feb 2017    
  
  
%\cite{Gursoy:2008bu}
\bibitem{Gursoy:2008bu} 
  U.~Gursoy, E.~Kiritsis, L.~Mazzanti and F.~Nitti,
  %``Deconfinement and Gluon Plasma Dynamics in Improved Holographic QCD,''
  Phys.\ Rev.\ Lett.\  {\bf 101}, 181601 (2008)
  doi:10.1103/PhysRevLett.101.181601
  [arXiv:0804.0899 [hep-th]].
  %%CITATION = doi:10.1103/PhysRevLett.101.181601;%%
  %150 citations counted in INSPIRE as of 06 Feb 2017  
  
  
  
 %\cite{Gursoy:2008za}
\bibitem{Gursoy:2008za} 
  U.~Gursoy, E.~Kiritsis, L.~Mazzanti and F.~Nitti,
  %``Holography and Thermodynamics of 5D Dilaton-gravity,''
  JHEP {\bf 0905}, 033 (2009)
  doi:10.1088/1126-6708/2009/05/033
  [arXiv:0812.0792 [hep-th]].
  %%CITATION = doi:10.1088/1126-6708/2009/05/033;%%
  
  
%\cite{Gubser:2008ny}
\bibitem{Gubser:2008ny} 
  S.~S.~Gubser and A.~Nellore,
  %``Mimicking the QCD equation of state with a dual black hole,''
  Phys.\ Rev.\ D {\bf 78}, 086007 (2008)
  doi:10.1103/PhysRevD.78.086007
  [arXiv:0804.0434 [hep-th]].
  %%CITATION = doi:10.1103/PhysRevD.78.086007;%%
  %152 citations counted in INSPIRE as of 06 Feb 2017
  %177 citations counted in INSPIRE as of 06 Feb 2017
 
  
%\cite{Gubser:2008yx}
\bibitem{Gubser:2008yx} 
  S.~S.~Gubser, A.~Nellore, S.~S.~Pufu and F.~D.~Rocha,
  %``Thermodynamics and bulk viscosity of approximate black hole duals to finite temperature quantum chromodynamics,''
  Phys.\ Rev.\ Lett.\  {\bf 101}, 131601 (2008)
  doi:10.1103/PhysRevLett.101.131601
  [arXiv:0804.1950 [hep-th]].
  %%CITATION = doi:10.1103/PhysRevLett.101.131601;%%
  %138 citations counted in INSPIRE as of 06 Feb 2017  

%\cite{Alanen:2009ej}
\bibitem{Alanen:2009ej} 
  J.~Alanen, K.~Kajantie and V.~Suur-Uski,
  %``Spatial string tension of finite temperature QCD matter in gauge/gravity duality,''
  Phys.\ Rev.\ D {\bf 80}, 075017 (2009)
  doi:10.1103/PhysRevD.80.075017
  [arXiv:0905.2032 [hep-ph]].
  %%CITATION = doi:10.1103/PhysRevD.80.075017;%%
  %27 citations counted in INSPIRE as of 15 Mar 2017

  %\cite{Alanen:2009xs}
\bibitem{Alanen:2009xs} 
  J.~Alanen, K.~Kajantie and V.~Suur-Uski,
  %``A gauge/gravity duality model for gauge theory thermodynamics,''
  Phys.\ Rev.\ D {\bf 80}, 126008 (2009)
  doi:10.1103/PhysRevD.80.126008
  [arXiv:0911.2114 [hep-ph]].
  %%CITATION = doi:10.1103/PhysRevD.80.126008;%%
  %37 citations counted in INSPIRE as of 15 Mar 2017


%\cite{Alanen:2009na}
\bibitem{Alanen:2009na} 
  J.~Alanen and K.~Kajantie,
  %``Thermodynamics of a field theory with infrared fixed point from gauge/gravity duality,''
  Phys.\ Rev.\ D {\bf 81}, 046003 (2010)
  doi:10.1103/PhysRevD.81.046003
  [arXiv:0912.4128 [hep-ph]].
  %%CITATION = doi:10.1103/PhysRevD.81.046003;%%
  %22 citations counted in INSPIRE as of 15 Mar 2017
  
  
%\cite{Alanen:2010tg}
\bibitem{Alanen:2010tg} 
  J.~Alanen, K.~Kajantie and K.~Tuominen,
  %``Thermodynamics of Quasi Conformal Theories From Gauge/Gravity Duality,''
  Phys.\ Rev.\ D {\bf 82}, 055024 (2010)
  doi:10.1103/PhysRevD.82.055024
  [arXiv:1003.5499 [hep-ph]].
  %%CITATION = doi:10.1103/PhysRevD.82.055024;%%
  %28 citations counted in INSPIRE as of 15 Mar 2017
  
  
  
  
%\cite{Kajantie:2011nx}
\bibitem{Kajantie:2011nx} 
  K.~Kajantie, M.~Krssak, M.~Vepsalainen and A.~Vuorinen,
  %``Frequency and wave number dependence of the shear correlator in strongly coupled hot Yang-Mills theory,''
  Phys.\ Rev.\ D {\bf 84}, 086004 (2011)
  doi:10.1103/PhysRevD.84.086004
  [arXiv:1104.5352 [hep-ph]].
  %%CITATION = doi:10.1103/PhysRevD.84.086004;%%
  %18 citations counted in INSPIRE as of 15 Mar 2017


%\cite{Alanen:2011hh}
\bibitem{Alanen:2011hh} 
  J.~Alanen, T.~Alho, K.~Kajantie and K.~Tuominen,
  %``Mass spectrum and thermodynamics of quasi-conformal gauge theories from gauge/gravity duality,''
  Phys.\ Rev.\ D {\bf 84}, 086007 (2011)
  doi:10.1103/PhysRevD.84.086007
  [arXiv:1107.3362 [hep-th]].
  %%CITATION = doi:10.1103/PhysRevD.84.086007;%%
  %17 citations counted in INSPIRE as of 15 Mar 2017
  
  
  
  %\cite{Kajantie:2013gab}
\bibitem{Kajantie:2013gab} 
  K.~Kajantie, M.~Krssak and A.~Vuorinen,
  %``Energy momentum tensor correlators in hot Yang-Mills theory: holography confronts lattice and perturbation theory,''
  JHEP {\bf 1305}, 140 (2013)
  doi:10.1007/JHEP05(2013)140
  [arXiv:1302.1432 [hep-ph]].
  %%CITATION = doi:10.1007/JHEP05(2013)140;%%
  %10 citations counted in INSPIRE as of 15 Mar 2017
  

%\cite{Li:2011hp}
\bibitem{Li:2011hp} 
  D.~Li, S.~He, M.~Huang and Q.~S.~Yan,
  %``Thermodynamics of deformed AdS$_5$ model with a positive/negative quadratic correction in graviton-dilaton system,''
  JHEP {\bf 1109}, 041 (2011)
  doi:10.1007/JHEP09(2011)041
  [arXiv:1103.5389 [hep-th]].
  %%CITATION = doi:10.1007/JHEP09(2011)041;%%
  %39 citations counted in INSPIRE as of 06 Feb 2017  

%\cite{Boyd:1996bx}
\bibitem{Boyd:1996bx} 
  G.~Boyd, J.~Engels, F.~Karsch, E.~Laermann, C.~Legeland, M.~Lutgemeier and B.~Petersson,
  %``Thermodynamics of SU(3) lattice gauge theory,''
  Nucl.\ Phys.\ B {\bf 469}, 419 (1996)
  doi:10.1016/0550-3213(96)00170-8
  [hep-lat/9602007].
  %%CITATION = doi:10.1016/0550-3213(96)00170-8;%%
  %906 citations counted in INSPIRE as of 07 Feb 2017


%\cite{Arnold:1995bh}
\bibitem{Arnold:1995bh} 
  P.~B.~Arnold and L.~G.~Yaffe,
  %``The NonAbelian Debye screening length beyond leading order,''
  Phys.\ Rev.\ D {\bf 52}, 7208 (1995)
  doi:10.1103/PhysRevD.52.7208
  [hep-ph/9508280].


%\cite{Bak:2007fk}
\bibitem{Bak:2007fk} 
  D.~Bak, A.~Karch and L.~G.~Yaffe,
  %``Debye screening in strongly coupled N=4 supersymmetric Yang-Mills plasma,''
  JHEP {\bf 0708}, 049 (2007)
  doi:10.1088/1126-6708/2007/08/049
  [arXiv:0705.0994 [hep-th]].


%\cite{Finazzo:2014zga}
\bibitem{Finazzo:2014zga} 
  S.~I.~Finazzo and J.~Noronha,
  %``Debye screening mass near deconfinement from holography,''
  Phys.\ Rev.\ D {\bf 90}, no. 11, 115028 (2014)
  doi:10.1103/PhysRevD.90.115028
  [arXiv:1411.4330 [hep-th]].
  %%CITATION = doi:10.1103/PhysRevD.90.115028;%%

%\cite{Finazzo:2015tta}
\bibitem{Finazzo:2015tta} 
  S.~I.~Finazzo,
  ``Understanding strongly coupled non-Abelian plasmas using the gauge/gravity duality,'' Phd Thesis, Universidade de S\~ao Paulo, 2015,  DOI 10.11606/T.43.2015.tde-07042015-144444 .  

%\cite{Iqbal:2008by}
\bibitem{Iqbal:2008by} 
  N.~Iqbal and H.~Liu,
  %``Universality of the hydrodynamic limit in AdS/CFT and the membrane paradigm,''
  Phys.\ Rev.\ D {\bf 79}, 025023 (2009)
  doi:10.1103/PhysRevD.79.025023
  [arXiv:0809.3808 [hep-th]].
  %%CITATION = doi:10.1103/PhysRevD.79.025023;%%
  %380 citations counted in INSPIRE as of 07 Feb 2017

%\cite{Gursoy:2012bt}
\bibitem{Gursoy:2012bt} 
  U.~Gürsoy, I.~Iatrakis, E.~Kiritsis, F.~Nitti and A.~O'Bannon,
  %``The Chern-Simons Diffusion Rate in Improved Holographic QCD,''
  JHEP {\bf 1302}, 119 (2013)
  doi:10.1007/JHEP02(2013)119
  [arXiv:1212.3894 [hep-th]].

%\cite{Grigoryan:2010pj}
\bibitem{Grigoryan:2010pj} 
  H.~R.~Grigoryan, P.~M.~Hohler and M.~A.~Stephanov,
  %``Towards the Gravity Dual of Quarkonium in the Strongly Coupled QCD Plasma,''
  Phys.\ Rev.\ D {\bf 82}, 026005 (2010)
  doi:10.1103/PhysRevD.82.026005
  [arXiv:1003.1138 [hep-ph]].
  %%CITATION = doi:10.1103/PhysRevD.82.026005;%%
  %33 citations counted in INSPIRE as of 17 Mar 2017

%\cite{Braga:2016wkm}
\bibitem{Braga:2016wkm} 
  N.~R.~F.~Braga, M.~A.~Martin Contreras and S.~Diles,
  %``Holographic Picture of Heavy Vector Meson Melting,''
  Eur.\ Phys.\ J.\ C {\bf 76}, no. 11, 598 (2016)
  doi:10.1140/epjc/s10052-016-4447-4
  [arXiv:1604.08296 [hep-ph]].
  

%\cite{Morningstar:1999rf}
\bibitem{Morningstar:1999rf} 
  C.~J.~Morningstar and M.~J.~Peardon,
  %``The Glueball spectrum from an anisotropic lattice study,''
  Phys.\ Rev.\ D {\bf 60}, 034509 (1999)
  doi:10.1103/PhysRevD.60.034509
  [hep-lat/9901004].
  %%CITATION = doi:10.1103/PhysRevD.60.034509;%%
  %749 citations counted in INSPIRE as of 07 Feb 2017



%\cite{Chen:2005mg}
\bibitem{Chen:2005mg} 
  Y.~Chen {\it et al.},
  %``Glueball spectrum and matrix elements on anisotropic lattices,''
  Phys.\ Rev.\ D {\bf 73}, 014516 (2006)
  doi:10.1103/PhysRevD.73.014516
  [hep-lat/0510074].
  %%CITATION = doi:10.1103/PhysRevD.73.014516;%%
  %349 citations counted in INSPIRE as of 07 Feb 2017

%\cite{Hoyos:2011uh}
\bibitem{Hoyos:2011uh} 
  C.~Hoyos, S.~Paik and L.~G.~Yaffe,
  %``Screening in strongly coupled N=2* supersymmetric Yang-Mills plasma,''
  JHEP {\bf 1110}, 062 (2011)
  doi:10.1007/JHEP10(2011)062
  [arXiv:1108.2053 [hep-th]].

%\cite{Singh:2012xj}
\bibitem{Singh:2012xj} 
  A.~Singh and A.~Sinha,
  %``Quantum corrections to screening at strong coupling,''
  Nucl.\ Phys.\ B {\bf 864}, 167 (2012)
  doi:10.1016/j.nuclphysb.2012.06.013
  [arXiv:1204.1817 [hep-th]].

%\cite{Andreev:2016hxm}
\bibitem{Andreev:2016hxm} 
  O.~Andreev,
  %``Color screening masses from string models,''
  Phys.\ Rev.\ D {\bf 94}, no. 12, 126003 (2016)
  doi:10.1103/PhysRevD.94.126003
  [arXiv:1608.08026 [hep-ph]].  
  
  
  
\end{thebibliography}
 \end{document}